\documentclass[pre,showpacs,twocolumn,amsmath,amssymb,10pt]{revtex4}
\newcommand*{\rom}[1]{\expandafter\@slowromancap\romannumeral #1@}
\usepackage[english]{babel}
\usepackage{amsmath}
\usepackage{graphicx}
\usepackage{dcolumn}
\begin{document}

\title{Energy barriers, entropy barriers, and non-Arrhenius behavior
in a minimal glassy model}

\author{Xin Du}
\author{Eric R. Weeks}
\affiliation{Department of Physics, Emory University, Atlanta,
Georgia 30322, USA}

\date{\today}

\begin{abstract}
We study glassy dynamics using a simulation of three soft Brownian
particles confined to a two-dimensional circular region.
If the circular region is large, the disks freely rearrange, but
rearrangements are rarer for smaller system sizes.  We directly
measure a one-dimensional free energy landscape characterizing
the dynamics.  This landscape has two local minima corresponding to
the two distinct disk configurations, separated by a free energy
barrier which governs the rearrangement rate.  We study several
different interaction potentials and demonstrate that the free
energy barrier is composed of a potential energy barrier and an
entropic barrier.  The heights of both of these barriers depend
on temperature and system size, demonstrating how non-Arrhenius
behavior can arise close to the glass transition.
\end{abstract}



\maketitle

\section{Introduction}

Glassy materials are amorphous solids:  disordered
microscopically, and unable to flow macroscopically
\cite{biroli13,ediger12,cavagna09,dyre06}.  They are inherently
out of equilibrium \cite{angell00,cipelletti02}, in contrast to
crystals.  In 1969, Goldstein proposed the idea of the potential
energy landscape, a conceptual framework for thinking about glassy
and crystalline materials \cite{goldstein69}.  The potential
energy landscape is defined as the potential energy $U$ of a
material ``plotted as a function of $3N$ atomic coordinates
in a $3N+1$ dimensional space,'' where $N$ is the number of
atoms \cite{goldstein69}.  At low temperatures, an ideal crystalline
solid will have particle coordinates that 
correspond to a global minimum of the potential energy landscape.
Glasses are disordered, so at low temperatures a glass will
have coordinates in a local minimum of the potential energy
landscape, but there are an enormous number of such local minima
\cite{debenedetti01,stillinger95,sciortino05rev,heuer08}.

Turning to higher temperatures where a material is a liquid,
thermal energy allows the system to rearrange constantly, and so
the $3N$ atomic coordinates trace out a trajectory traversing the
potential energy landscape.  If the temperature is close to the
material's glass transition, and if crystallization is avoided,
then the trajectory through the landscape spends most of its
time near local minima, with occasional passages through a saddle
point in the landscape to an adjacent minimum
\cite{stillinger82,sastry98}.
The number of minima, their depth, and the details of the saddles
between them can be connected to the microscopic
dynamics of samples at a variety of temperatures
\cite{sciortino05rev,heuer08}.
At low temperatures, the thermal energy $k_B
T$ does not allow the system to escape a local minimum easily.
In particular the escape from any particular local minimum is
a thermally activated process, depending on the barrier height
between that local minimum and the minima adjacent in the $3N+1$
dimensional space.  Of course, given the high dimensionality of
the problem, visualizing this is impossible except for conceptual
sketches \cite{stillinger95,perezcastaneda14,charbonneau14},
of which the earliest one we are aware of was by Stillinger and
Weber in 1984 \cite{stillinger84}.

The picture of a potential energy landscape changes when
one considers a system of hard spheres.  Hard spheres are
defined as particles that have no interaction energy when
they are not in contact, and infinite interaction energy if
they touch.  As a function of the $3N$ sphere coordinates, the
potential energy surface is flat at $U=0$ except for prohibited
configurations for which $U=\infty$.  Rather than local minima
separated by saddles, the landscape has flat open areas
separated by bottlenecks that correspond to {\it entropic}
barriers.  As hard spheres can form glasses at high densities
\cite{cohen59,woodcock81,speedy98}, these entropic barriers must
function similarly to the potential energy barriers in a potential
energy landscape \cite{bowles99,ashwin09,hunter12pre}.

\begin{figure}[bt]
\centering
\includegraphics[scale=0.4]{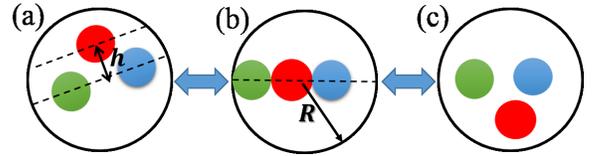}
\caption{\label{sketch} (Color online) Sketch of our model, with
three distinguishable particles confined within a circular system.
In (a), $h$ is the distance between one of the particles and the
axis defined by the other two. In (b), $R$ is the radius of the
confining boundary.  The states (a)-(c) show a cage breaking event in
our model, where $h$ changes sign.}
\end{figure}

In 2012 Hunter and Weeks introduced a simple model with hard
particles where the entropic landscape was directly measurable
\cite{hunter12pre}.  The model consists of three hard disks executing
Brownian motion within a two-dimensional circular region.  
As illustrated in Fig.~\ref{sketch},
the system has two distinct configurations of the three disks.
A transition occurs between these two configurations when any one
of the three particles passes between the other two.  When the
system size $R$ is smaller, these transitions are rarer.
This model captures the flavor of hard spheres near their
glass transition, where rearrangements are difficult due to
particle crowding \cite{perera99,doliwa00}.  Hunter and Weeks
directly calculated a free energy landscape based entirely on the
entropy of the states.  They demonstrated that the transition time
scale was related to the entropic barrier height,
$\tau \sim \exp(S_b)$.

In the current manuscript, we extend the model of Hunter and
Weeks to consider the case of soft particles.  In this situation,
we now have a potential energy landscape that varies smoothly as a
function of the particle coordinates.  However, the best description
of our model is through the free energy landscape which includes
both entropy and potential energy.  The transition state shown
in Fig.~\ref{sketch}(b) still corresponds to a barrier, now with
both potential energy and entropic components.  For finite-range
interaction potentials, we use simulations to see how the free energy
landscape approaches the hard disk case as $T \rightarrow 0$.
For all potentials, we examine potential energy and entropy to
understand the relative importance of each in determining the
transition rate between states.  Our most significant result is an
explicit demonstration that the influences of both potential energy
and entropy depend on temperature; that is, the effective free
energy barrier height depends on $T$.  Our results help bridge concepts
between soft and hard particles in a simple model, complementing
prior molecular dynamics simulations done with large numbers of
soft particles \cite{berthier09,xu09,schmiedeberg11}.

Our model is a straightforward system with non-Arrhenius scaling
as the glass transition is approached.  Arrhenius scaling
occurs in a system where a time scale $\tau$ for a transition
is set by a fixed energy barrier of size $\Delta$, such that
$\tau \sim \exp(-\Delta/k_B T)$.  In a glass-forming system,
$\tau$ could be the time scale for diffusion or flow, and $\tau$
grows dramatically as the glass transition is approached.  Often,
this happens in a non-Arrhenius fashion \cite{angell95}: $\tau$
grows faster than expected as $T$ is decreased.  This leads to the
interpretation that $\Delta = \Delta(T)$ increases as $T$ decreases.
We demonstrate that in our model $\Delta$ is due to potential
energy and entropy, both of which are $T$-dependent, even though
the underlying potential energy landscape is $T$-independent.

\section{The Model System}
\label{potentials}

We study three two-dimensional particles confined to a circular
system of size $R$ as shown in Fig.~\ref{sketch}.  We will consider
four distinct particle interactions in our simple model system.

Our first particle type is a commonly used finite-ranged harmonic
potential \cite{durian95,ohern03}.  This considers deformable soft particles
interacting through purely repulsive body centered forces.
Our harmonic potential is defined as:
\begin{equation}
U_{\mathrm{HM}}(r_{ij}) = 
\left\{\begin{array}{lc} 
U_{0}(\frac{2 - r_{ij}}{2})^{2}; & r_{ij} < 2 \\
0;& r_{ij} \geq 2\end{array}\right.
\label{hm}
\end{equation}
Here \(r_{ij}\) is the center-to-center distance between
particles \(i\) and \(j\).  All particles have radius 1 ($a_i =
a_j = 1$) and do not interact when they are not touching ($r_{ij}
\geq 2$).  
The particles have the same interaction with the wall: 
\begin{equation}
U_{\mathrm{HM, wall}}(r_{ic}) = \left\{
\begin{array}{lc}
U_{0}(\frac{ r_{ic}-(R-1)}{2})^{2}; & r_{ic} > R-1 \\
0;& r_{ic} \leq R-1 \end{array}
\right.
\label{hmwall}
\end{equation}
$r_{ic}$ is the distance between the particle center and system
center, that is, it is the radial coordinate of particle $i$.  As
the particle radius is 1, when $r_{ic} = R - 1$ the particle
comes into contact with the wall, and for $r_{ic} > R-1$, the
interaction energy increases and the particle feels a repulsive
force from the wall.

Our second particle type is also repulsive, but has a infinite range
interaction between the particles, and between the particles and
the wall; we term this the ``long-range potential.''  We define
this potential as:
\begin{equation}
U_{\mathrm{LR}}(r_{ij}) = U_0(\frac{r_{ij}}{2})^{-12}
\label{lr}
\end{equation}
between the particles and
\begin{equation}
U_{\mathrm{LR, wall}}(r_{ic}) = U_0(\frac{R-r_{ic}}{2})^{-12}
\end{equation}
between the particles and the wall.

Our third particle type uses the Lennard-Jones potential (``LJ
potential''), which
approximates the interaction between a pair of neutral atoms
\cite{lennardjones31}.
The Lennard-Jones potential is defined as:
\begin{equation}
U_{\mathrm{LJ}}(r_{ij}) =
U_0(\frac{r_{ij}}{2})^{-12}-U_0(\frac{r_{ij}}{2})^{-6}.
\label{lj}
\end{equation}
This interaction potential differs from the first two (harmonic
and long-range) in that Lennard-Jones particles have both a
repulsive and an attractive component.  In contrast to the
first two potentials, these particles have a finite preferred
separation distance that minimizes $U$ at $r_{ij} = 2^{7/6} =
2.245$.  To simplify this model, the wall is hard. In this case,
the interaction energy with the wall is $U=0$ until the particles
touch the wall ($r_{ic} = R-1$) in which case $U=\infty$.

We consider one last particle type using the
Weeks-Chandler-Andersen potential (``WCA potential'')
\cite{andersen71}.  This potential starts with the LJ
potential, truncates it at the minimum, and then shifts it upward
so that the potential goes smoothly to zero:
\begin{equation}
U_{\mathrm{WCA}}(r_{ij}) = \left\{\begin{array}{lc}
U_{LJ}+\frac{U_0}{4}; & r_{ij} <2^{7/6} \\
0; & r_{ij} \geq 2^{7/6}\end{array}\right.
\label{wca}
\end{equation}
This then is the repulsive component of the LJ potential,
and has no attractive component.  As with the LJ
potential, we again assume a strictly hard contact with the
confining wall.  Like the harmonic potential, the WCA potential is
finite-ranged, but in contrast this potential diverges at $r_{ij}
\rightarrow 0$.  This latter behavior is like the long-range
potential, which also diverges.

These four interaction potentials capture several interesting
possibilities.  Two are finite-ranged; three are purely
repulsive; three diverge as the particle separation goes to zero.

We use the Metropolis algorithm \cite{metropolis53} to simulate
Brownian motion of the particles, similar to previous work by our
group \cite{hunter12pre}.  At each Monte Carlo step, we try to move
each particle (one at a time) in a random direction with $rms$ step
size of 0.01 (or in some cases smaller).  We consider the change
in energy $\Delta U$ for the trial move.  These trial moves are
accepted with probability 1 if $\Delta U < 0$, and with probability
$\exp(-\beta \Delta U)$ otherwise, with $\beta = 1/k_B T$.
The initial condition is with the three particles starting at the
vertices of an equilateral triangle of side length 2, and the
system is equilibrated after the first transition of the sort
shown in Fig.~\ref{sketch}.  Each simulation was run for at least
20 transitions (in the cases with very slow dynamics) and more
typically 100-1000 transitions.  Given that there is no memory in
this system, each condition was run only once as a time-average
was adequate (although we did check this with multiple runs
several times, and also checked that the results are insensitive
to the $rms$ step size).

In all situations, the radius of the confined system is $R$
as indicated in Fig.~\ref{sketch}.  For the harmonic
potential, recall the particle radius is 1, so for $R=3$ the
particles can just line up across a diameter of the system with
$U=0$.  For $R < 3$, particles can only change configuration
[Fig.~\ref{sketch}(a-c)] with a nonzero temperature.  The WCA
potential is also finite-ranged, although the range is not 1 but
rather $2^{7/6}$, so here $R=
1+2^{\frac{7}{6}} = 3.245$ is the smallest radius at which
particles can line up across a diameter with $U=0$.  
For the long-range potential and the LJ
potential, particles always interact with nonzero potential
energy, and so there is no value of $R$ with any special meaning.

Note that the meaning of $U_0$ differs between the potentials in
an unimportant way.  For the harmonic potential, $U_0$ is the
maximum potential energy between two particles when they are
fully overlapped ($r_{ij}=0$).  For the long-range potential,
$U_0$ is the potential energy between two particles when 
$r_{ij} = 2$.  For LJ and WCA, there are yet other
meanings for $U_0$.  In the simulation, we simply set $U_0=1$
and vary the value of $k_B T$.  As $U_0$
is not comparable between the different interactions, likewise
specific values of $T$ are not comparable either.  Accordingly,
our discussion will focus on comparing behaviors as functions of
$T$ without need to compare specific values.  The remainder of
the paper will study the behavior of our model as we change $R$,
$T$, and the interaction potential.  In particular we are most
interested as the system becomes ``glassy:'' smaller $R$ and/or
smaller $T$.


\section{Results}

\subsection{Free energy landscapes}

To study the free energy landscape, we define a macrostate variable
$h$ as is shown in Fig.~\ref{sketch}(a) \cite{hunter12pre}.
To do this, we pick two particles to define an axis (say,
pointing from particle 1 to particle 2).
$h$ is the distance of the third particle above (or
below) this axis.  $h$ can be positive or negative, and is zero
at the transition state shown in Fig.~\ref{sketch}(b). Therefore,
when $h$ changes sign, a rearrangement occurs.  It is arbitrary
which particles are used to define the horizontal axis; if we
consider $h'$ and $h''$ defined using different pairs of
particles, all three $h$ variables change sign simultaneously
upon a transition \cite{hunter12pre}.

Following Ref.~\cite{hunter12pre}, we construct the free energy
landscape by counting occurrences of each $h$ in the simulation
for given parameters ($R$ and $T$).  We then compute $P(h)$,
the probability of seeing each $h$ value.  Finally,
the free energy landscape is computed directly according to the
Boltzmann distribution, $P(h) \sim \exp(-F(h)/k_BT)$. For simplicity,
we set \(k_B=1\) in the simulation.  We shift $F(h)$ so that the
minimum value is $F=0$.

\begin{figure}
\centering
\includegraphics[width=0.48\textwidth]{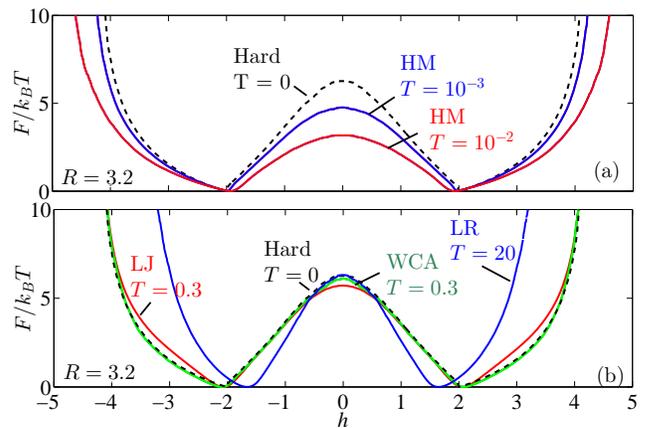}
\caption{\label{landscape} (Color online) (a) The free energy landscape
for the harmonic (``HM'') interaction potential.  $R=3.2$ and the
temperatures are as indicated.
(b) Free energy landscapes for interactions as
indicated, where ``LR'' designates the long-range potential,
``LJ'' the Lennard-Jones potential, and ``WCA'' the
Weeks-Chandler-Andersen potential.  $R=3.2$ and the temperatures
are as indicated, chosen so that the barrier height at $h=0$ is
comparable for the different interaction potentials.
}
\end{figure}

Figure \ref{landscape}(a) shows the free energy landscape for
the harmonic potential model.  There is a free
energy barrier at the transition state $h=0$.  For $R=3.2$, the
particles do not have to overlap at the transition state, but for
$T>0$ they are allowed to overlap which makes transitions easier.
Keeping $R$ fixed, as $T \rightarrow 0$ overlaps are less likely,
and the free energy barrier for transitions grows.  At $T=0$,
overlaps are impossible, although since this is a finite-range
potential, transitions still occur.  In this situation the free
energy landscape is identical to the landscape for hard disks,
indicated by the dashed line in Fig.~\ref{landscape}(a).  In other
words, at low \(T\), thermal fluctuations decrease and
these soft particles become hard.

The other features of the free energy landscape shown in
Fig.~\ref{landscape}(a) are straightforward to understand.
There are two symmetrically located minima close to $h = \pm 2$
that correspond to the most probable states for the three particles
\cite{hunter12pre}.  For large values of $|h|$, the particles
are forced to interact with the confining wall.  This causes
the free energy to grow dramatically due to the large potential
energy penalty.

Figure \ref{landscape}(b) shows free energy landscapes for other
interaction potentials, with temperatures chosen so that the
barrier height is approximately the same for each, and $R=3.2$
kept constant.  The shapes are all qualitatively similar, although
the long range potential has particles confined to a smaller range
of $h$.  For the hard particle case, the minima occur precisely
at $h = \pm 2$ \cite{hunter12pre}.  For the other potentials,
the locations of the minima vary with $T$.  For the LR and LJ
potentials, one can compute the configuration that minimizes $U$,
and the $h$ that minimize $F(h)$ are fairly close to the $h$ for
those minimal $U$ configurations.  The $T$ dependence, however,
makes it clear that minimizing the free energy is not the same as
minimizing the potential energy.  Maximizing entropy plays
a role as well in determining the $h$ that minimizes $F(h)$.
As previously reported, in the hard model, $\partial F / \partial
h$ is discontinuous at $h = \pm 2$ \cite{hunter12pre}. However,
this derivative is continuous everywhere in all of the soft models.

\subsection{Dynamics and free energy barriers}

\begin{figure}
\centering
\includegraphics[width=0.48\textwidth]{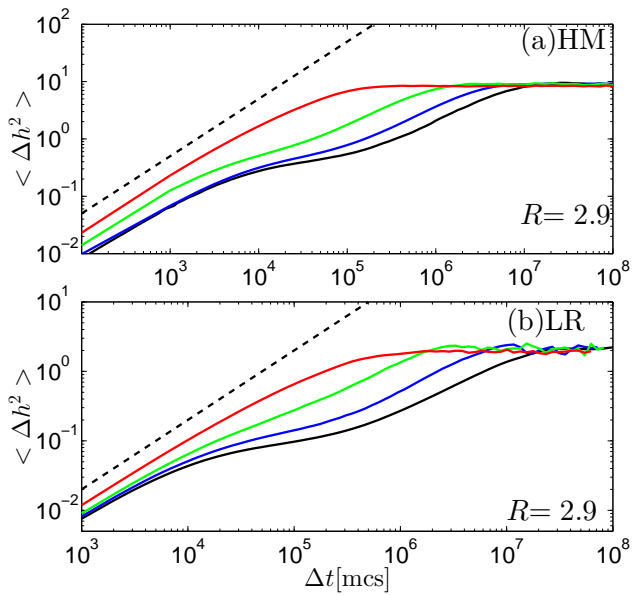}
\caption{\label{msdhm} (Color online) (a) Mean-square displacement 
in $h$ space for the harmonic (``HM'') interaction potential. $R=2.9$ and $T = 10^{-1},10^{-2},10^{-2.2},10^{-2.4}$ (from top to bottom, red to black). The dashed line has a slope of 1. (b) Mean-square displacement in $h$ space for the long-range (``LR'') interaction potential.  $R=2.9$ and $T = 10^{6},10^{4.6},10^{4.2},10^{4}$ (from top to bottom, red to black). The dashed line has a slope of 1.}\end{figure}

The dynamics are straightforward when considering $h(t)$.  Often,
$h(t)$
stays close to the values $h_{\rm min}$ that minimize the free
energy landscape (Fig.~\ref{landscape}), but occasionally $h(t)$
switches sign.  We quantify the dynamics by plotting the mean
square displacement (MSD) $\langle \Delta h^2 \rangle$ as a function of
lag time $\Delta t$ in Fig.~\ref{msdhm} for the harmonic
potential (a) and long-range potential (b).  At the shortest
times, particles diffuse fairly freely.  At intermediate time
scales, the MSD starts to level off, reflecting
that the system is trapped in one of the probable states shown
in Fig.~\ref{sketch}(a,c).  At longer time scales, the system can
swap between these two states, and the MSD
begins to rise again.
At the longest time scales shown in Fig.~\ref{msdhm}, the MSD
levels off due to the finite system size.

\begin{figure}
\centering
\includegraphics[width=0.48\textwidth]{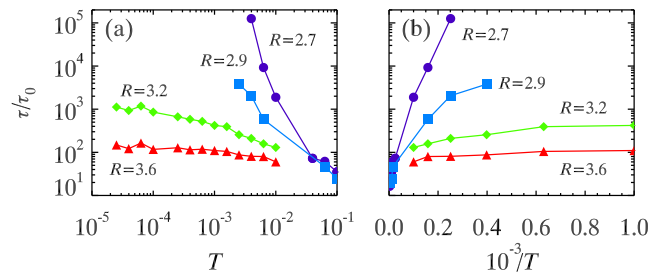}
\caption{\label{nonarr}(Color online) Dependence of
$\tau$ on $T$ and $R$ in harmonic potential system. Curves in
different colors show the life time as a function of
$T$ with different $R$ as indicated.
The lifetimes $\tau$ are 
normalized by $\tau_0 = 1/2D$, the time a free particle would
take on average to diffuse a distance of 1, using the diffusion
constant $D$ from the simulation.
}
\end{figure}

To quantify the transition time scale $\tau$, we measure the
average time between sign changes of $h$.  However, during a
transition, there are often small fluctuations right around $h=0$
that are not true transitions.  To avoid biasing $\tau$ toward
lower time scales, we stipulate that once $h = 0$ is crossed,
the system must move a further distance \(\Delta{h}=1\) before
returning \cite{hunter12pre}; our results are not sensitive
to this choice.  The probability distribution of time scales
$P(\tau)$ is exponentially distributed so the mean value gives the
appropriate time scale, which we plot in Fig.~\ref{nonarr} as a
function of temperature (a) and inverse temperature (b).  The two
largest system sizes $R$ show a horizontal leveling off of $\tau$
at cold temperatures.  This is the limit where the soft particles
behave as hard particles, and $\tau$ reaches the value seen for
purely hard particles \cite{hunter12pre}.  For the smaller system
sizes, particles must overlap to have a transition, and so as $T
\rightarrow 0$ this becomes rare and $\tau$ diverges.  Were any
of these systems to be Arrhenius with a temperature-independent
potential energy barrier, the data in Fig.~\ref{nonarr}(b) would
fall on a straight line; that they do not indicates that the system
is non-Arrhenius.

\begin{figure}
\centering
\includegraphics[width=0.47\textwidth]{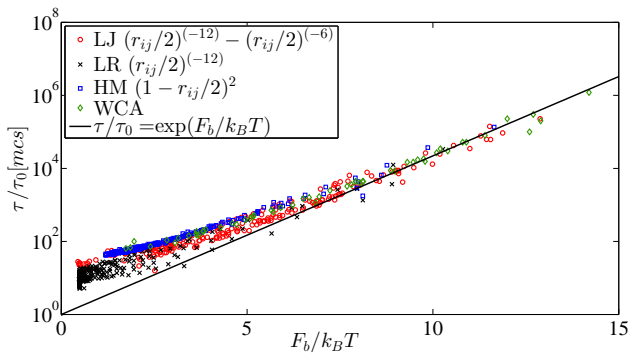}
\caption{\label{arrhenius}The lifetime grows exponentially with
the free energy barrier $F_b$ as $F_b \rightarrow \infty$, as
indicated by the solid line.  
The symbols are as indicated, and correspond to a variety of $R$
and $T$ values; for specific ranges of $R$ and $T$, see ranges
shown in Figs.~\ref{nonarr}, \ref{fbarriers}, and
\ref{entropyenergy}.
}
\end{figure}

An alternate way to think of Arrhenius behavior is in terms of the
free energy barrier for transitions, $F_b$.  Calculating the free
energy landscapes as in Fig.~\ref{landscape} allows us to determine
$F_b = F(h=0)$.  Transitions are less frequent with higher $F_b$.
Figure~\ref{arrhenius} verifies that $\tau$ grows Arrheniusly as
a function of $F_{b}$, $\tau \sim \exp(\beta F_b)$ as $F_b
\rightarrow \infty$.  The deviations seen for small $F_b$ are due
to large system sizes:  for larger systems, it simply takes longer
for particles to move to the transition state \cite{hunter12pre}.
The details of this vary depending on the potential.  Additionally,
the vertical spread of symbols for a given potential for $F_b
\lesssim 5$ reflects that different $R$ and $T$ values can have
the same $F_b$.  Nonetheless, the collapse of the data at larger
$F_b$ indicates that $\tau$ grows Arrheniusly with $F_b$
precisely where the dynamics are slowest.

\begin{figure}
\centering
\includegraphics[width=0.48\textwidth]{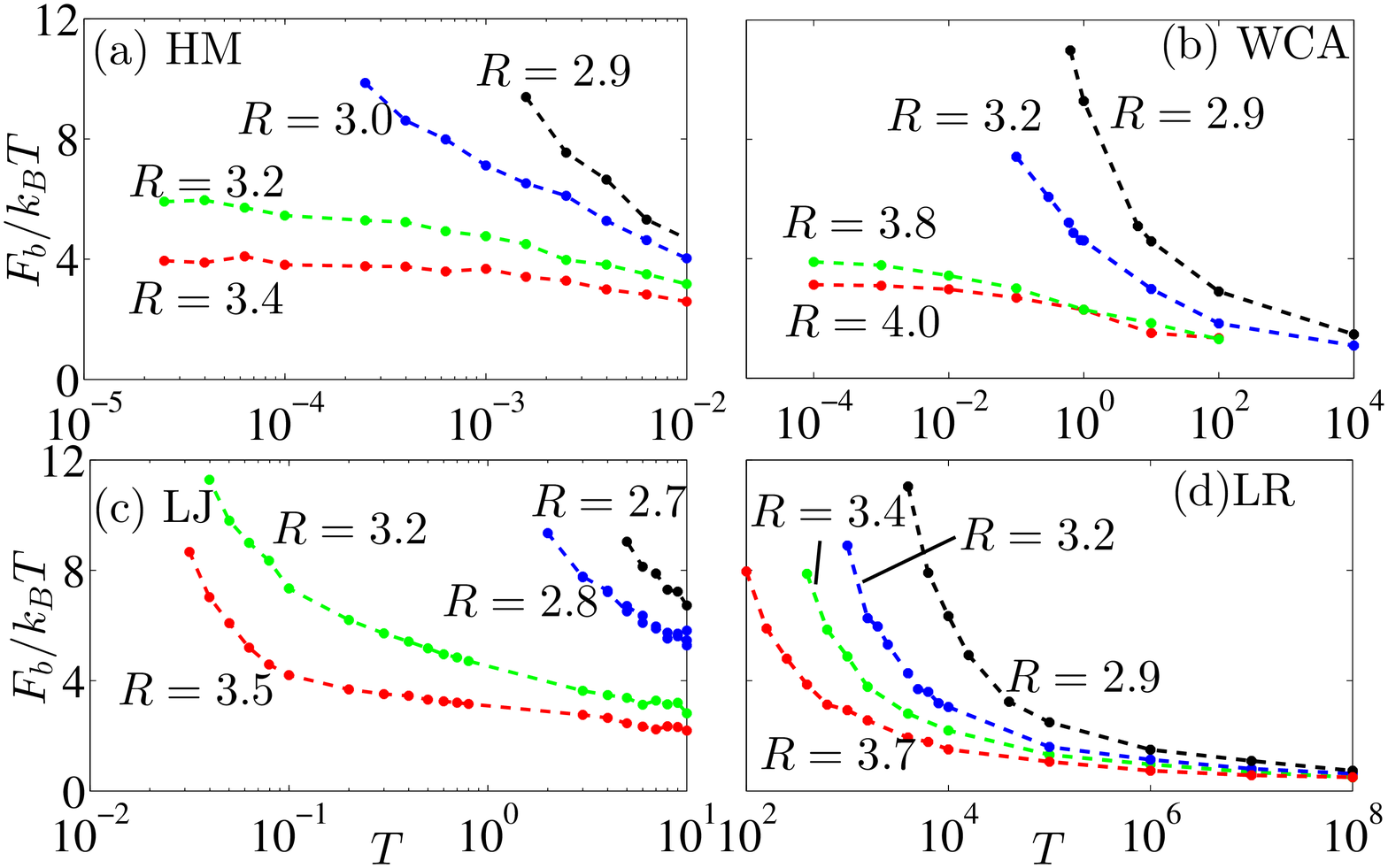}
\caption{\label{fbarriers}(Color online) Dependence of
\(\frac{F_{b}}{k_BT}\) on $T$ and $R$. Curves in
different colors show the free energy barrier as a function of
$T$ with different $R$ as indicated.  The interaction potentials
are (a) harmonic (``HM''), (b) WCA, (c) Lennard-Jones, (d)
long-range (``LR'').
(As discussed in Sec.~\ref{potentials}, recall that the
specific values of $T$ are not comparable between the different
potentials.)
}
\end{figure}

Our primary interest is understanding the cause of glassy
dynamics in our system.  In other words, we'd like to understand
how $\tau$ grows large (equivalently, how $F_b$ grows large) as
we decrease $T$ and/or decrease $R$.  Figure \ref{fbarriers} shows
$F_b/k_B T$ as a function of $T$ for different particle types.  In each
panel, the different curves are for different system sizes $R$.
As expected, $F_b/k_B T$ grows with decreasing $T$ and with decreasing
$R$.  Panels (a) and (b) show some curves with qualitatively
different behavior, in that $F_b/k_B T$ goes to a plateau as $T
\rightarrow 0$.  As with Fig.~\ref{nonarr}, this
is because of the behavior of the free
energy landscape shown in Fig.~\ref{landscape}(a) for these two
finite-ranged potentials:  for large system sizes $R$, even at $T=0$
the particles can rearrange without overlapping.  For large $R$,
the plateau values for $F_b$ seen in Fig.~\ref{fbarriers}(a,b)
are precisely the free energy barrier heights for hard disks
\cite{hunter12pre}.  For this argument to work, the system size $R$
must exceed a critical value, $R_c = 3$ for the harmonic potential
and $R_c = 3.245$ for the WCA potential (as discussed at the end
of Sec.~\ref{potentials}).  For $R<R_c$, particles must overlap
at $h=0$ with $U>0$, and so as $T \rightarrow 0$ the free energy
barrier $F_b$ will diverge.  For the LJ and LR potentials, at $h=0$
we always have $U>0$ and so not surprisingly $F_b$ diverges in all
cases at low temperatures, with the details depending
on $R$.  

These behaviors raise an interesting question.  In the cases of
Fig.~\ref{fbarriers}(a,b) with a plateau, the system approaches
the hard disk behavior as $T \rightarrow 0$.  For hard disks,
this free energy barrier is entirely an entropic barrier
\cite{hunter12pre}.  However, clearly for many
other cases in Fig.~\ref{fbarriers}, the free energy barrier is
at least in part due to the potential energy component of the
barrier.  To what extent in any of these cases can the free
energy barrier be ascribed to entropy, and to what extent to
potential energy?

\subsection{Simple models for the transition state}
\label{models}

\begin{figure}
\centering
\includegraphics[width=0.48\textwidth]{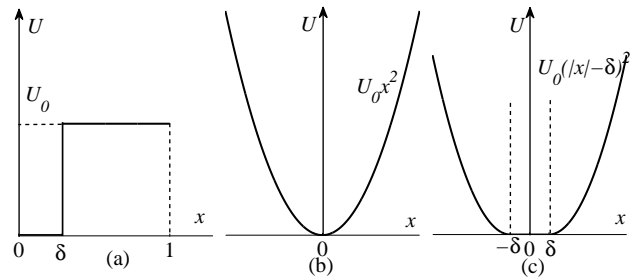}
\caption{\label{model}Sketch of three simple potential energy
landscapes.  (a) Model 1.
(b) Model 2. (c) Model 3.}
\end{figure}

To understand the interplay of entropy and potential energy at the
transition state ($h=0$) for our three particle system, we consider
a simple model for the transition state.  Consider a system moving
along a reaction coordinate $h$ with a flat energy landscape, except
for a barrier at $h=0$.  At $h=0$, we will assume there is a second
coordinate $x$ in the orthogonal direction. In the three particle
system, this would account for other degrees of freedom for the
particle locations subject to the constraint $h=0$.  We examine
three ideas for $U(x)$, sketched in Fig.~\ref{model}.

First consider Model 1 [Fig.~\ref{model}(a)], where we let $x$
be constrained on the interval $0 \leq x \leq 1$ and the potential
energy barrier depends on $x$ as:
\begin{eqnarray}
U(x) = 0, \ \ 0 \leq x < \delta \\
U(x) = U_0, \ \ \delta \leq x \leq 1
\label{model1}
\end{eqnarray}

\noindent so that the system can either make a transition at
zero potential energy cost, or with a finite cost $U_0 > 0$.

Attempts to cross with zero potential energy cost occur with
probability
\begin{equation}
p_1=\delta
\end{equation}
and these attempts always succeed.  Attempts to cross elsewhere
occur with probability $(1-\delta)$ and succeed with probability
$\exp(-U_0/k_B T)$; thus the likelihood of a barrier crossing taking this
pathway is 
\begin{equation}
p_2 = (1-\delta) \exp(-U_0/k_B T).
\end{equation}
The crossing attempt entirely fails with probability $1-p_1-p_2$.
If attempts are made with a time scale $\tau_0$, then the mean
transition time can be shown to be

\begin{equation}
\tau = \frac{\tau_0}{p_1 + p_2}.
\end{equation}

\noindent The question to consider, then, is what this transition
looks like in terms of a free energy barrier, if we average
over the coordinate $x$?  Two limits are immediately obvious.
If $U_0/k_B T$ is sufficiently large, $p_1 \gg p_2$ and the
transition rate is governed by an entropic barrier.  In the
converse limit, if $\delta$ is sufficiently small, the $U=0$
pathway is vanishingly rare ($p_1 \ll p_2$) and transitions are governed by the
potential energy barrier $U_0$. In between these limits, one can think
of this system as having an effective free energy barrier that
is due to both potential energy and entropy.  The mean potential
energy the system has when the barrier is crossed is given by

\begin{equation}
\beta \langle U \rangle = \frac{\beta U_0 p_2}{p_1 + p_2}
\label{modelenergy}
\end{equation}

\noindent using $\beta = 1/k_B T$.  The partition function at
the crossing is given by $Z = p_1 + p_2$, the free energy barrier
height is $\beta F = -\ln Z = -\ln (p_1 + p_2)$, and the entropy
can be derived as

\begin{equation}
\label{modelentropy}
\beta T S = -\beta T \frac{\partial F}{\partial T} = \ln(p_1 + p_2) +
\beta \langle U \rangle
\end{equation}

\noindent (which is also apparent from $F = U - TS$).

The conclusion is that while the potential energy surface
is $T$-independent and always has a $U=0$ transition pathway,
the free energy barrier depends on $T$ and on average requires
nonzero potential energy for the transition.  Given that $p_2$
depends on $T$, Eqns.~\ref{modelenergy} and
\ref{modelentropy} show that both the potential energy and entropy
contributions to the free energy barrier depend on $T$.

We next consider the more realistic Model 2, where the transition
has a harmonic potential with respect to the coordinate $x$:
\begin{equation}
U(x) = U_0 x^2.
\label{model2}
\end{equation}
For this potential, the mean potential energy required is $\beta
\langle U \rangle = 1/2$ (equipartition).  In the interesting limit
$T \rightarrow 0$, the free energy barrier grows as $\beta F \sim |
\ln T |$.  As the potential energy contribution is independent of
$T$, the barrier growth is due to entropy:  at low temperatures the
system only crosses at $|x| \lesssim \sqrt{k_B T/U_0}$.  As with
Model 1, while $U(x)$ is independent of $T$, the free energy
barrier depends on $T$.

Finally, we consider model 3 which is a hybrid of the previous
two models:
\begin{eqnarray}
U(x) = 0, \ \  |x| < \delta \\
U(x) = U_0(|x| - \delta)^2, \ \  |x| \geq\delta.
\label{model3}
\end{eqnarray}
In this model, the mean potential energy required
to cross the barrier is $ \beta \langle U \rangle =
\frac{1}{2}({1+2\delta\sqrt[]{\frac{\beta U_0}{\pi}}})^{-1}$. At
low temperature and with large $\delta$, the system prefers to
cross within the region $ |x| < \delta$ where potential energy
is zero.  In this case, $\delta\sqrt[]{\frac{\beta U_0}{\pi}}
\rightarrow \infty$, and $ \beta \langle U \rangle\rightarrow 0$.
For small $\delta$ and/or large $T$, the average potential energy
found when crossing the barrier is larger.  At high temperature
and with small $\delta$, when $\delta\sqrt[]{\frac{\beta U_0}{\pi}}
\rightarrow 0$ $, \beta \langle U \rangle \rightarrow 1/2$, which
is same as model 2.

To be clear, for these models we are really interested in the
case where the system climbs a potential energy hill to reach the
transition state $h=0$.  We are then considering how the system
crosses through the $h=0$ state, and concluding that this requires
additional potential energy (on average) and also navigating an
entropic barrier.  In other words, merely having enough potential
energy to reach the saddle is insufficient, as threading through
the saddle's lowest point is of low probability. In all of
these simple models of the transition state, the transition
time scale will be
\begin{equation}
\tau = \tau_0 \exp(\beta U_{\rm min}) \exp(\beta F)
\end{equation}
where $U_{\rm min}$ is the potential energy of the saddle's lowest
point, and $F$ is the additional free energy barrier associated with
the $h=0$ potential energy landscape cross-section. The $\exp(\beta
U_{\rm min})$ contribution gives Arrhenius scaling with $T$, and
the $\exp(\beta F)$ contribution provides additional non-Arrhenius
scaling as $F$ grows with decreasing $T$, as shown above.  In
many situations, the $\exp(\beta U_{\rm min})$ term dominates,
but it depends on the details as will be shown below.

\subsection{Barriers:  Energy and Entropy}

\begin{figure}
\centering
\includegraphics[width=0.48\textwidth]{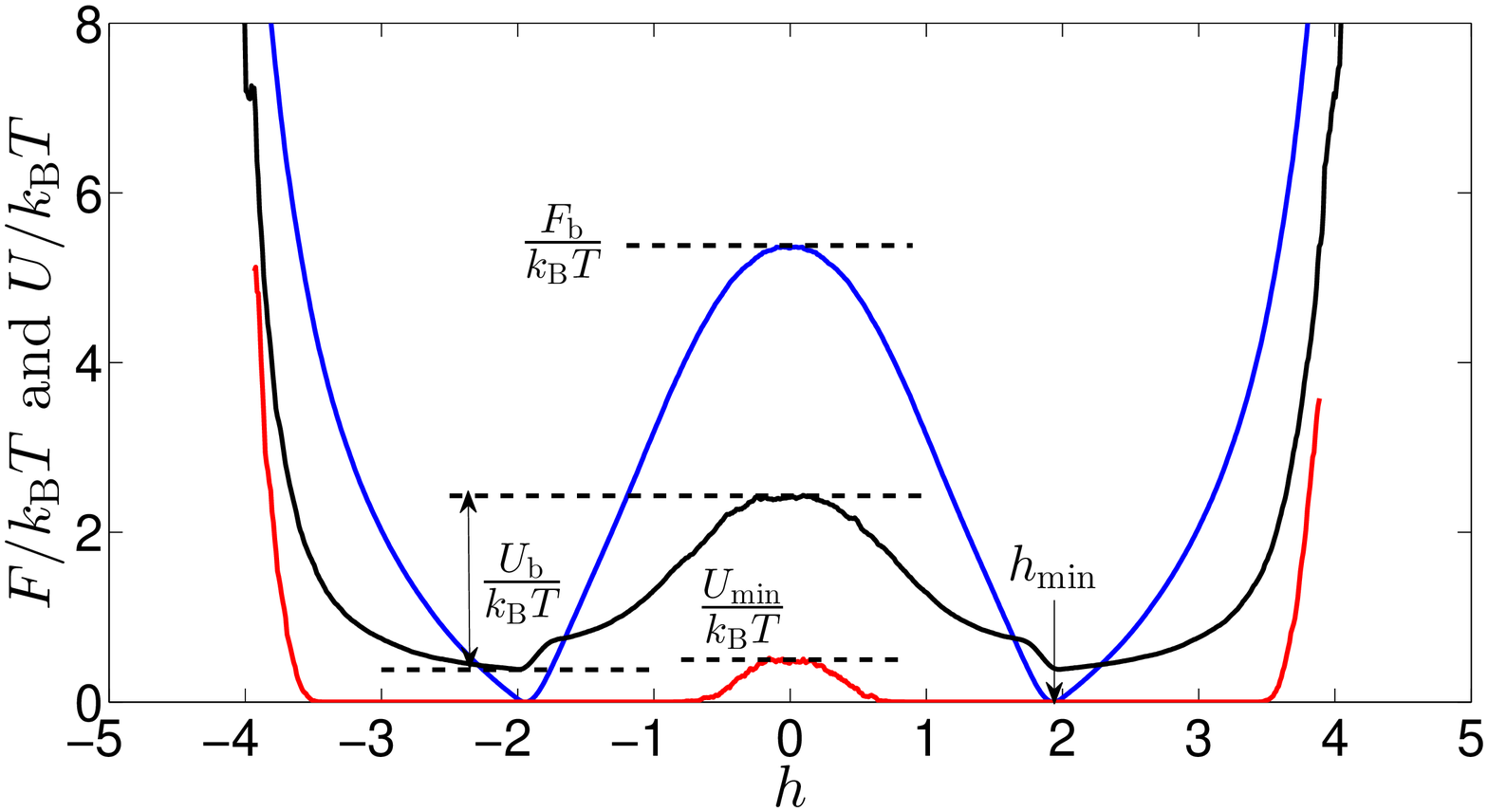}
\caption{\label{ufmin}(Color online) The free energy landscape
for the harmonic (``HM'') interaction potential.  $R=2.9$ and
$T=10^{-2.2}$.The blue curve is the free energy landscape. The black
curve is the potential energy landscape based on $\langle U \rangle$
in $h$ space. The red curve is the minimum potential energy,
$U_{\rm min}$, in $h$ space. $F_b/k_BT$, $U_b/k_BT$, and $U_{\rm min}/k_BT$
are as indicated.  $h_{\rm min}$ is the value that minimizes $F$.
}
\end{figure}

This discussion motivates us to divide the free energy barrier in
our three-particle simulations into energetic and entropic
components.  As $F = U - TS$, we consider the free energy barrier
to be:
\begin{equation}
\beta F_b = \beta U_b + S_{b}
\label{barriercomponents}
\end{equation}
where as usual, $k_B=1$.  The relevant quantities are illustrated
in Fig.~\ref{ufmin}.  $h_{\rm min}$ is the value of $h$ that
minimizes the free energy.  The contribution of potential energy
to the barrier is defined as $U_b = \langle U \rangle(0) -
\langle U\rangle (h_{\rm min})$.  $\langle U \rangle$ is the
black curve
indicated by $U_b$ in Fig.~\ref{ufmin}, and is averaged over
$\geq 20$ barrier crossings.
Equation \ref{barriercomponents} lets us calculate $S_b$ from
$F_b$ and $U_b$.  Note that the definition of $S_b$ differs from
$U_b$ by a minus sign:
$S_b = S(h_{\rm min}) - S(h) > 0$, such that it
is positive (and thus a barrier).  
The minimum possible potential energy for
each value of $h$ is the thin red curve in Fig.~\ref{ufmin} which
is at zero for most values of $h$.  We define $U_{\rm min}$ as the
minimum potential energy needed to cross $h=0$ if the system finds
the optimum transition path, as indicated in Fig.~\ref{ufmin}.
It is clear from Fig.~\ref{ufmin} that $U_b$ will almost always
be larger than $U_{\rm min}$, although a rare exception for the
Lennard-Jones potential will be described below.  $U_{\rm min}$ is a
quantity we can derive analytically for each interaction potential,
while $U_b$ is determined from the simulation data.  $U_{\rm min}$
is temperature independent, in contrast to $U_b$, $S_b$, and $F_b$.
We wish to see what conditions allow $S_b$ or $U_b$ to dominate
the free energy barrier, and also to gain some intuition about
non-Arrhenius temperature dependence in general.  Note that
simulation times become nearly intractable when $\beta F_b =
\beta U_b + S_b \gtrsim 10$, thus limiting how much of the growth
of the barriers we can study.

\begin{figure}
\centering
\includegraphics[width=0.48\textwidth]{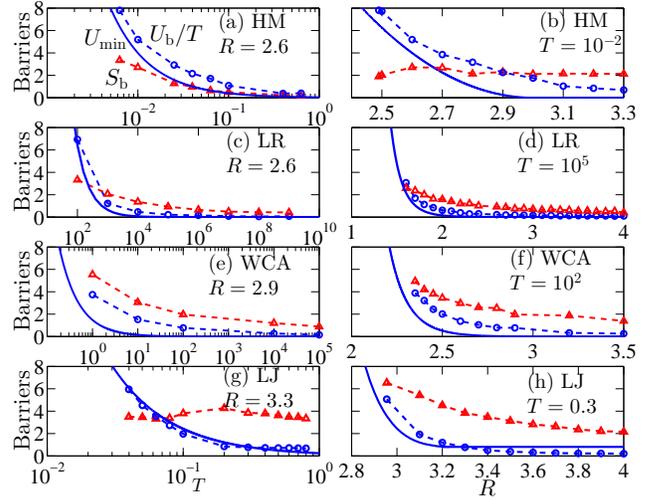}
\caption{\label{entropyenergy} (Color online) Data for the potential energy
barrier and entropy barrier for a variety of interaction
potentials, temperatures, and system sizes $R$, as indicated.
The solid curves indicate $U_{\rm min}$, the theoretical minimum
potential energy barrier.  The symbols indicate measured values
(from the simulation data),
connected by straight dashed lines.  The meanings of the curves
are all as labeled in (a).  The interaction potentials
are (a,b) harmonic (`HM'), (c,d) long-range (`LR'),(e,f)
Weeks-Chandler-Andersen (`WCA') and (g,h) Lennard-Jones (`LJ').
}
\end{figure}

Figure \ref{entropyenergy}(a) shows data for the harmonic
interaction potential for $R = 2.6$.  As $R < 3$, the particles
must overlap at $h=0$ and thus $U_{\rm min} > 0$.  The graph
shows that as $T \rightarrow 0$, both $\beta U_b$ and $S_b$ grow.
The growth of $\beta U_b$ is more significant, pushed up by $\beta
U_{\rm min}$.  This situation is analogous to Model 2 from
Sec.~\ref{models}, where $U_b \approx U_{\rm min} + \frac{1}{2}k_B
T$ and $S_b \approx | \ln T |$.  Given the slow growth of $S_b$,
the free energy barrier is dominated by $U_{\rm min}$.

Figure \ref{entropyenergy}(b) shows complementary data for the
harmonic interaction potential as a function of $R$ at fixed $T =
10^{-2}$.  Given the finite range for the potential, $U_{\rm min}
= 0$ for $R \geq 3$, although $U_b > 0$ as the particles 
overlap for some crossings.  The $R>3$ case is analogous to
Model 3, whereas $R<3$ is analogous to Model 2.  The data in
Fig.~\ref{entropyenergy}(b) show that entropy plays a smaller
role for small $R$, where the free energy barrier is dominated by
the potential energy.  For this interaction potential, 
$U_{\rm min} \sim (R-3)^2$ for $R<3$; the data show that
$S_b$ is nearly constant as a function of $R$.

Figure \ref{entropyenergy}(c,d) shows the comparisons of entropic
barrier and potential barrier for the LR potential.  For large $T$
or large $R$ cases, $S_b > \beta U_b$.  In the converse cases,
the opposite is true.  As the system becomes slower with a large
free energy barrier, the free energy barrier is strongly determined
by the potential energy component.

For the WCA data shown in Fig.~\ref{entropyenergy}(e,f), $U_{\rm
min}$ goes to zero at $R=3.245$, although as before we still have
$\beta U_b > 0$.  For the WCA potential, we see a more dramatic
growth of $U_b$ with decreasing $T$ [panel (e)] and with
decreasing $R$ [panel (f)].  It appears that if we further shrink
the system size in panel (f), $U_b$ will eventually grow larger
than $S_b$.  The growth of $S_b$ at small $R$ is not as strong as
the growth of $\beta U_{\rm min}$, and since $U_b > U_{\rm
min}$, this further suggests that $\beta U_b$ will be larger than
$S_b$ for smaller systems.

Figure \ref{entropyenergy}(g,h) shows the comparisons of entropic
barrier and potential barrier for the LJ potential.  For 
high $T$ cases [panel (g)], $S_b > \beta U_b$, with the opposite
occurring as $T \rightarrow 0$.  Panel (h) shows that at a fixed
$T$, with decreasing $R$ both $S_b$ and $\beta U_b$ grow, with
the latter growing more dramatically.
It appears that if we further
shrink the system size in panel (h), $U_b$ will eventually grow
larger than $S_b$.

Unusual behavior is seen for the LJ potential in
Fig.~\ref{entropyenergy}(g,h), where $U_b < U_{\rm min}$
with large $R$ and low $T$.  This can be understood given the
differences between our definitions of $U_b$ and $U_{\rm min}$.
$U_{min}$ considers the difference in potential energy between
the lowest potential energy path at the saddle point ($h=0$) and
the lowest potential energy the particles can obtain given $R$.
The latter corresponds to a configuration where the centers of the
particles form an equilateral triangle with side length $=2.24$,
corresponding to $h=1.94$.  However, this configuration is itself
an unlikely configuration, and for example when $h=1.94$ the
three particles will often be in a configuration with slightly
higher potential energy than the absolute minimum.  This is
essentially the same argument put forth in Sec.~\ref{models},
that the average potential energy experienced by the system is
not the minimum value.  Thus, the {\it measured} potential energy
difference $U_b$ will often be between a slightly higher value for
both $h=0$ and $h=h_{\rm min}$, such that their difference $U_b
= U(0) - U(h_{\rm min}) < U_{\rm min}$.  This is not the
case for the other interaction potentials, probably because the
potential energy is a flatter function of $h$ around $h_{\rm min}$
for the other interaction potentials.

Some general conclusions can be drawn from all of the data of
Fig.~\ref{entropyenergy}.  First, in most of the cases, $U_b >
U_{\rm min}$, confirming the intuition from Sec.~\ref{models}:
that crossing the saddle point in the potential energy landscape is
not typically done at the minimal potential energy path through
that saddle point.  Second, Fig.~\ref{entropyenergy}(a,c,e)
demonstrates that $\beta U_b$ and $S_b$ both depend on $T$ and are
larger for colder temperatures:  and thus these barriers behave
non-Arrheniusly.  In particular, these barriers are not simply
based on $\beta U_{\rm min}$.

\begin{figure}
\centering
\includegraphics[width=0.48\textwidth]{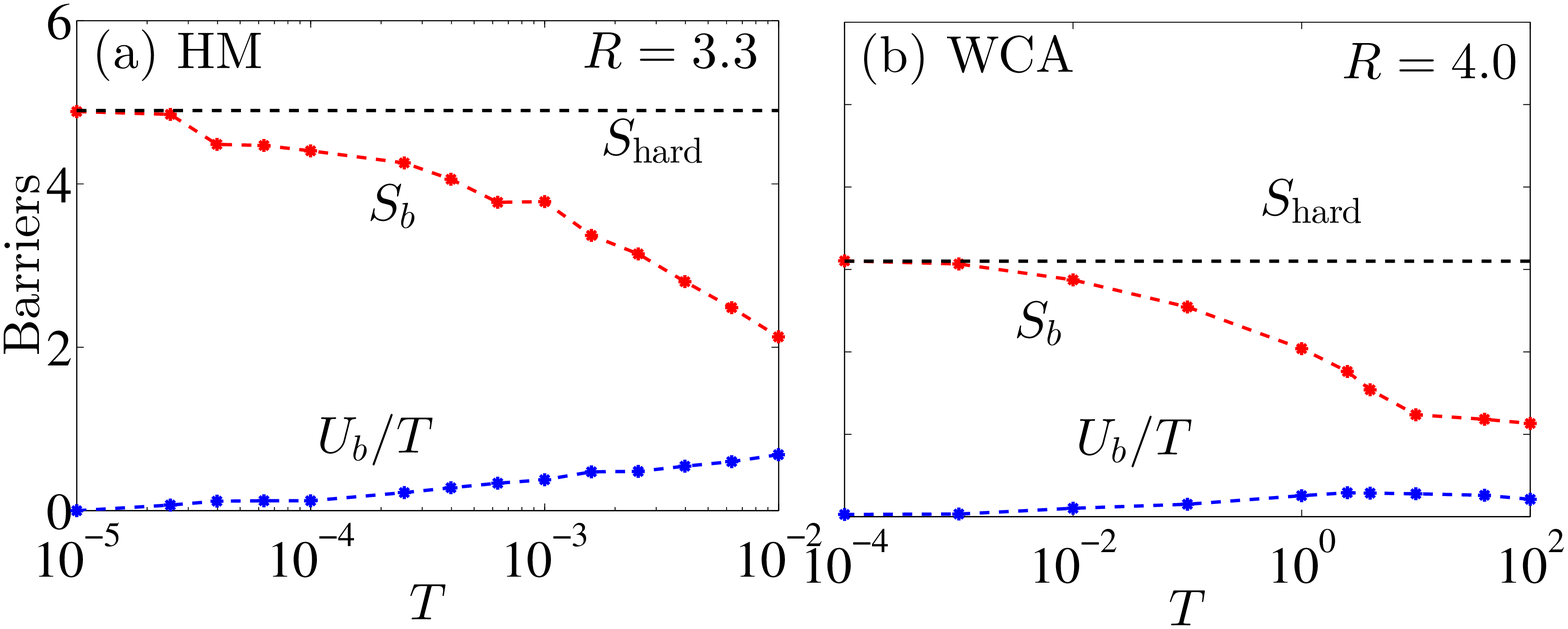}
\caption{\label{posepsilon} (Color online) The potential barrier and entropy
barrier for (a) the harmonic interaction potential and (b) the WCA
interaction potential.  The system sizes $R$ are as indicated,
and chosen such that the minimum potential energy barrier is
$U_{\rm min} = 0$.  The horizontal dashed lines indicate the free
energy barrier for the hard disk case.
}
\end{figure}

The finite-ranged potentials (harmonic and WCA) allow us to look
at cases where $U_{\rm min} = 0$.  As noted in the discussion of
Fig.~\ref{fbarriers}(a,b), when $U_{\rm min}=0$ the free energy
barriers reach a plateau as $T \rightarrow 0$ corresponding to
the hard disk limit (the horizontal dashed lines).  The data
for the energy and entropy barriers are shown for two of these
cases in Fig.~\ref{posepsilon}.  These data match the qualitative
behavior predicted by Model 3 (Sec.~\ref{models}).  At low $T$,
$\beta U_b \approx 0$ and $S_b$ approaches the hard disk result.
At high $T$, $\beta U_b \approx \frac{1}{2}$ and the entropic
contribution decreases as more microstates are possible at $h=0$.
For different temperatures, the trade-off between crossing with
zero or finite potential energy changes, due to the entropic
penalty of choosing the zero potential energy pathway, which is
weighted by the temperature.

\section{Conclusions}

We studied a free energy landscape of a simple model possessing
some qualitative features of a glass transition.  The model's slow
dynamics are governed by a free energy barrier which we directly
measure in simulations.  The barrier height is determined both
by entropy and potential energy.  The relative contributions of
each of these depend on temperature $T$.  In particular, for fixed
system size $R$, the potential energy landscape is independent of
$T$, yet the effective potential energy barrier height, entropic
barrier height, and overall free energy barrier all depend on $T$.
This leads to non-Arrhenius temperature dependence.
We conjecture that for cases with more particles, entropy plays
an even more important role in cooperative rearrangements, as
suggested in 1965 by Adam and Gibbs \cite{adam65} and discussed
by many authors subsequently.  In fact, our model is quite in
the spirit of Adam and Gibbs, in that rearrangements require
coordinated motion of all three particles [Fig.~\ref{sketch}(b)]
which results in an entropic penalty.

There are qualitative differences between our model results and
non-Arrhenius behavior seen in glass-forming systems.  First, the
onset of slow dynamics in our model requires temperature changes
of several orders of magnitude (Fig.~\ref{fbarriers}), whereas
similar changes in glassy materials require a temperature decrease
of only 10-20\% \cite{biroli13,ediger12,cavagna09,dyre06,angell95}.
Second, in our model, as $T \rightarrow 0$, the potential energy
component of the barrier may become more important than entropy,
suggesting a possible recovery of Arrhenius behavior at the lowest
$T$ (Fig.~\ref{entropyenergy}).  However, this is not completely
clear from our data as the $T\rightarrow 0$ behavior requires
prohibitively long simulation runs.  Both of these differences
between our simple model and glassy behavior might disappear for
larger numbers of particles, but then we would lose the ability to
fully visualize the free energy landscape (Fig.~\ref{landscape}).
It is certainly known that near the glass transition,
rearrangements can involve far more than three particles
\cite{perera99,doliwa00}, which would likely enhance the
sensitivity to temperature.
While we do not provide a realistic description of the $N
\rightarrow \infty$ limit of a glass transition, we have
demonstrated connections between the free energy landscape, free
energy barriers, and non-Arrhenius temperature dependence in our
model glassy system.

We thank M.~E.~Cates, F.~Family and G.~L. Hunter for helpful
discussions. This work has been supported financially by the NSF
(CMMI-1250199/-1250235).

\bibliography{threedisk}

\end{document}